\documentclass[12pt]{article}
\usepackage{graphicx}
\usepackage{multirow}
\newcommand\pubdate{\today}

\def\pisa{Universit\`a di Pisa and INFN - Sezione di Pisa\\ giulia.casarosa@pi.infn.it\\\vspace{2mm}(on behalf of the \babar\ Collaboration)}

\def\Title#1{\begin{center} {\Large #1 } \end{center}}
\def\Author#1{\begin{center}{ \sc #1} \end{center}}
\def\Address#1{\begin{center}{ \it #1} \end{center}}

\newcommand\pubblock{\rightline{\begin{tabular}{l} \\%\pubnumber\\
         \pubdate  \end{tabular}}}
\newenvironment{Abstract}{\begin{quotation}  }{\end{quotation}}
\newenvironment{Presented}{\begin{quotation} \begin{center} 
             PRESENTED AT\end{center}\bigskip 
      \begin{center}\begin{large}}{\end{large}\end{center} \end{quotation}}
\def\Acknowledgements{\bigskip  \bigskip \begin{center} \begin{large}
             \bf ACKNOWLEDGEMENTS \end{large}\end{center}}
\RequirePackage{xspace}
\usepackage{relsize}
\def\babar{\mbox{\slshape B\kern-0.1em{\smaller A}\kern-0.1em
    B\kern-0.1em{\smaller A\kern-0.2em R}}}

\def\yCP        {\ensuremath{y_{C\!P}}\xspace}
\def\deltaY     {\ensuremath{\Delta Y}\xspace}

\def\tauhhp     {\ensuremath{\tau^+}\xspace}
\def\tauhhm     {\ensuremath{\bar{\tau}^+}\xspace}
\def\tauKpi     {\ensuremath{\tau_{K\pi}}\xspace}
\def\Dz      {\ensuremath{D^0}\xspace}
\def\Dbar    {\kern 0.2em\overline{\kern -0.2em D}{}\xspace}
\def\Dzb     {\ensuremath{\Dbar^0}\xspace}
\def\DzDzb         {\ensuremath{\Dz\ensuremath{-}\Dzb}\xspace}
\def\CP                {\ensuremath{C\!P}\xspace}
\def\CPV        {\ensuremath{C\!P\!V}\xspace}
\def\CPT               {\ensuremath{C\!PT}\xspace} % Looks better without \!
\def\Kpm   {\ensuremath{K^\pm}\xspace}
\def\pimp  {\ensuremath{\pi^\mp}\xspace}
\def\DzRW       {\ensuremath{\Kpm \pimp}\xspace}
\def\dm         {\ensuremath{\Delta m}\xspace}
\def\terr       {\ensuremath{\sigma_{t}}\xspace}
\def\Pchisq      {\ensuremath{P(\chi^2)}}
\def\Kp    {\ensuremath{K^+}\xspace}
\def\Km    {\ensuremath{K^-}\xspace}
\def\pip   {\ensuremath{\pi^+}\xspace}
\def\pim   {\ensuremath{\pi^-}\xspace}\def\KpKm  {\ensuremath{\Kp \kern -0.16em \Km}\xspace}
\def\pippim     {\ensuremath{\pi^{+}\pi^{-}}\xspace}
\def\Dstp        {\ensuremath{D^{*+}}}
\def\epem       {\ensuremath{e^+e^-}\xspace}
\def\invfb   {\ensuremath{\mbox{\,fb}^{-1}}\xspace}
\def\Y#1S{\ensuremath{\Upsilon{(#1S)}}\xspace}% no space before {...}!
\def\FourS {\Y4S}
\def\pisoftp    {\ensuremath{\pi_{\rm s}^{+}}\xspace}
\def\piz   {\ensuremath{\pi^0}\xspace}
\def\ps   {\ensuremath{\rm \,ps}\xspace}
\def\fs   {\ensuremath{\rm \,fs}\xspace}
\def\Dp      {\ensuremath{D^+}\xspace}
\def\kk         {\ensuremath{KK}\xspace}
\def\pep2{PEP-II}

\newcommand{\chisq}{\ensuremath{\chi^2}\xspace}
\newcommand{\gevc}{\ensuremath{{\mathrm{\,Ge\kern -0.1em V\!/}c}}\xspace}
\newcommand{\mevc}{\ensuremath{{\mathrm{\,Me\kern -0.1em V\!/}c}}\xspace}
\newcommand{\gevcc}{\ensuremath{{\mathrm{\,Ge\kern -0.1em V\!/}c^2}}\xspace}
\newcommand{\mevcc}{\ensuremath{{\mathrm{\,Me\kern -0.1em V\!/}c^2}}\xspace}
\newcommand{\stat}{\ensuremath{\mathrm{(stat)}}\xspace}
\newcommand{\syst}{\ensuremath{\mathrm{(syst)}}\xspace}
\newcommand{\gae}{\lower 2pt \hbox{$\, \buildrel {\scriptstyle >}\over {\scriptstyle \sim}\,$}}

\def\beq{\begin{equation}}
\def\eeq#1{\label{#1}\end{equation}}
\def\beqa{\begin{eqnarray}}
\def\eeqa#1{\label{#1}\end{eqnarray}}

\begin{document}

%\setpagewiselinenumbers % line numbers
%\modulolinenumbers[1]
%\linenumbers

\begin{titlepage}
\pubblock

\vfill
\Title{Measurement of $\DzDzb$ Mixing and \CP Violation at \babar}
\vfill
\Author{Giulia Casarosa}%\\(on behalf of the \babar\ Collaboration)}
\Address{\pisa}
\vfill
\begin{Abstract}
We report on a measurement of $\DzDzb$ mixing and a search for \CP violation in the
$\Dz\to\KpKm,\,\pippim \mbox{ and } \DzRW$ channels. We use \Dz's coming from \Dstp\ decays, so that
 the flavour of the \Dz at production is tagged by the charge of the pion that is also emitted. We also use an independent
set of \Dz's coming directly from the hadronization of the charm quark, but in this case the flavour of the charmed meson is not known.
We analyze events collected by the \babar\ experiment at the PEP-II
asymmetric-energy \epem collider, corresponding to an integrated luminosity of 468\invfb.
We measure the mixing parameter value to be $\yCP =  [0.72 \pm 0.18 \stat \pm 0.12 \syst]\%$, and exclude the no-mixing 
hypothesis at $3.3\sigma$ significance. We find no evidence of \CP violation, observing
$\deltaY  =  [0.09 \pm 0.26 \stat \pm 0.06 \syst]\%$  which is consistent with zero.
\end{Abstract}
\vfill
\begin{Presented}
%Charm 2012\\
The $5^{\rm th}$ International Workshop on Charm Physics\\
Honolulu, Hawai'i,  May 14--17, 2012
\end{Presented}
\vfill
\end{titlepage}
\def\thefootnote{\fnsymbol{footnote}}
\setcounter{footnote}{0}

\section{Introduction}

%Mixing and \CP violation (\CPV) measurements have been of great importance in the Standard Model (SM) development.
Mixing in the charm sector is a well-established phenomenon~\cite{Aubert:2007wf,Aubert:2007en,Aubert:2009ai,Staric:2007dt,Abe:2007rd,CDF:2007uc} although there is no single measurement that exceeds 
$5\sigma$ significance. Recently the LHCb~\cite{Aaij:2011in} and CDF~\cite{Collaboration:2012qw} Collaborations %Aaltonen:2011se
have reported evidence of \CP violation (\CPV) in the difference of the integrated \CP  asymmetries in the $\Dz\to\KpKm$ 
and $\Dz\to\pippim$ channels. This result was unexpected at the current experimental precision, and it may be a manifestation of New Physics (NP),
although a Standard Model (SM) explanation cannot be ruled out. These measurements have renewed the interest of the community in charm physics as a sector in which to search for NP manifestations.

Under the hypothesis of \CPT conservation the two mass eigenstates ($D_1$ and $D_2$) can be written in terms of the flavor eigenstates (\Dz and \Dzb) as:
\beq
| D_{1,2} \rangle = p | \Dz \rangle \pm q | \Dzb \rangle \qquad \mbox{with} \qquad  \left|p\right|^2 + \left|q\right|^2 = 1.
\eeq{eq:qpdef}
If $\CP | \Dz \rangle = +| \Dzb \rangle $, then in the case of no \CPV, $D_1$ is the $\CP$-even state and $D_2$ the $\CP$-odd state.
The parameters that describe \DzDzb oscillations are proportional to the difference
of masses ($m_i$) and widths ($\Gamma_i$) of the mass eigenstates:
\beq
x \equiv \frac{m_1 - m_2}{\Gamma} \qquad \mbox{ and } \qquad y \equiv \frac{\Gamma_1 - \Gamma_2}{2\Gamma},
\eeq{eq:xyDef}
where $\Gamma = (\Gamma_1+\Gamma_2)/2$ is the average width. 
Mixing will occur if the mass eigenstates differ from the flavour eigenstates, that is, if either $x$ or $y$ is non-zero.
SM predictions for the mixing parameter values are at the order of a percent or less and, at present, experimental measurements are
in agreement with these predictions. Unfortunately the theoretical predictions are affected by large computational uncertainties on the
dominant long-range-diagram contributions, preventing these measurements from being strong tests of the SM.

In the following we present a measurement of the mixing parameter \yCP~\cite{Liu:1994ea} and the \CP-violating parameter \deltaY, defined as:
\beq
\yCP \equiv \frac{\Gamma^++\bar{\Gamma}^+}{\Gamma}-1 \qquad \mbox{and} \qquad \deltaY \equiv \frac{\Gamma^+ - \bar{\Gamma}^+}{2\Gamma},
\eeq{eq:defycp}
where $\Gamma^+$  ($\bar{\Gamma}^+$) is the average width of the \Dz (\Dzb) when reconstructed in \CP-even eigenstates.

The measured values of \yCP and \deltaY constrain the parameters that govern mixing and \CPV\ in the charm sector.
Neglecting the effect of direct \CP violation, estimated to be at least one order of magnitude below our current sensitivity,
we relate \yCP and \deltaY to the mixing and \CP-violating parameters as follows:
\beqa
\yCP  = y\cos\phi - \frac{A_M}{2} x\sin\phi \qquad \mbox{ and } \qquad \deltaY = -x\sin\phi + \frac{A_M}{2} y\cos\phi.
\eeqa{eq:interms}
The asymmetry $A_M =  \frac{ (q/p)^2 - (p/q)^{2}}{(q/p)^2 + (p/q)^{2}}$ measures \CP violation in mixing, while $\phi$ is sensitive to \CPV in the interference between decays with and 
without mixing, being the weak phase of the quantity $\lambda_f = \frac{q}{p}\frac{\bar{A}_f}{A_f}$ with $A_f$ $(\bar{A}_f)$ the amplitude for the decay $\Dz (\Dzb) \to f$.
In principle, $\phi$ can depend also on the final state, but with our current level of precision we are not sensitive to this~\cite{Kagan:2009gb}.
In the absence of \CP violation $\yCP = y$ and $\deltaY = 0$.

\section{Data Sample and Backgrounds}

We reconstruct the \Dz in the $h^+h^-$ ($h = K, \pi$) and \DzRW final states and measure three lifetimes: 
\begin{itemize}
\item \tauhhp for the $\Dz \to  h^+h^-$ decays,
\item  \tauhhm for the $\Dzb \to  h^+h^-$ decays,
\item \tauKpi for the $\Dz$ (and \Dzb) $ \to  \DzRW$ decays (the Cabibbo favored $\Km\pip$ and the doubly Cabibbo suppressed $\Kp\pim$ decays are collected in the same sample).
\end{itemize}
Due to the small mixing rate ($\le 1\%$) we can neglect the effect of mixing and assume that all signal proper time distributions are exponential.
The untagged $\Dz\to\KpKm$ sample~\cite{CC:2008} is assumed to contain 50\% of \Dz and 50\% of \Dzb decays.
The three values of inverse lifetime are used to compute \yCP and \deltaY: \tauKpi is used to access the average width $\Gamma$ while,
\tauhhp (\tauhhm) is used to obtain $\Gamma^+$ ($\bar{\Gamma}^+$).

We use {\it tagged} decays of the \Dz %\footnote{Charge conjugation is implied throughout.},
 coming from \Dstp\ decays, through $\Dstp\to\Dz\pisoftp$, as well as {\it untagged} decays
coming directly from the hadronization of the charm quark. The tagged and untagged samples are independent, 
{\it i.e.} an event containing a tagged candidate and at least one untagged candidate is excluded from the untagged sample.
 In the tagged sample the flavour of the \Dz is determined by the charge of the pion that is also emitted.
Due to the significantly higher level of background in  the \pippim final state, we do not use the related untagged sample.

We analyze $468 \invfb$ of data recorded by the \babar\ detector~\cite{Aubert:2001tu} at, and slightly below, the $\FourS$ resonance 
at the  $\epem$ asymmetric-energy \hbox{PEP-II} $B$-Factory .
To avoid potential bias, we finalize our data selection criteria,
as well as the procedures for fitting, extracting statistical limits,
and determining systematic uncertainties, prior to examining the results.

An oppositely charged pair of $\Kp$ or $\pip$ candidates satisfying particle identification criteria
is fit to a common vertex to form a \Dz candidate.
We require each \Dz to have momentum in the center-of-mass (CM) frame $p_{\rm CM} > 2.5\gevc$ in order to remove almost completely
\Dz's coming from $B$-meson decays.
For the tagged modes, we form the \Dstp\ candidate by fitting a \Dz candidate and a charged pion track $\pisoftp$ to a common vertex, which is required
to lie within the \epem interaction region. The  $\pisoftp$ momentum is required to be greater than 0.1 \gevc in the laboratory frame and less than
0.45 \gevc in the CM frame. We veto any $\pisoftp$ candidate that may have originated from a reconstructed
photon conversion or \piz Dalitz decay and reject a positron that fakes a $\pisoftp$ candidate
by using energy loss information.
We also select tagged candidates in a \dm window, $0.1447 < \dm < 0.1463\gevcc$, where $\dm$ is the difference
between the reconstructed \Dstp\ and \Dz masses. This requirement strongly suppresses backgrounds.

The proper time $t$ and  proper time error $\terr$ of each \Dz candidate are
determined from a combined fit to the \Dz production and decay vertices.
The \chisq probability of the vertex fit must satisfy $\Pchisq > 0.1\%$.
We retain only candidates with $-2 < t < 4 \ps$
and $\terr < 0.5 \ps$.
For tagged decays, this fit  does not
incorporate any $\pisoftp$ information in order
to ensure that the lifetime resolution models for tagged and untagged
signal decays are very similar.
The most probable value of \terr for signal events
is $\sim 40\%$ of the nominal $\Dz$ lifetime~\cite{Nakamura:2010zzi}.

For cases where multiple $\Dstp$ candidates in an event
share one or more tracks and the \Dz decays to the same final state (\Km\pip and \Kp\pim are considered to be the same final
state in this context), we retain only the candidate with the highest
\Pchisq. If an event contains a tagged decay, all untagged candidates
from that event are excluded from the final sample. In an event with no $\Dstp$ candidate and
multiple $\Dz$ candidates decaying to the same final state, we retain only the $\Dz$ candidate with the highest
\Pchisq.
%If an event contains a tagged decay, all untagged candidates
%from that event are removed.
%For cases where multiple \Dz (for the untagged modes) or \Dstp (for the tagged modes) 
%candidates in an event share one or more tracks, we retain only the candidate with the highest
%\Pchisq. 
The fraction of events with multiple $\Dz$ candidates with overlapping daughter tracks is $\ll 1\%$ for all final states.

\begin{figure}[htb]
%\centering
\hbox to \hsize{
  \includegraphics[width=0.24\linewidth]{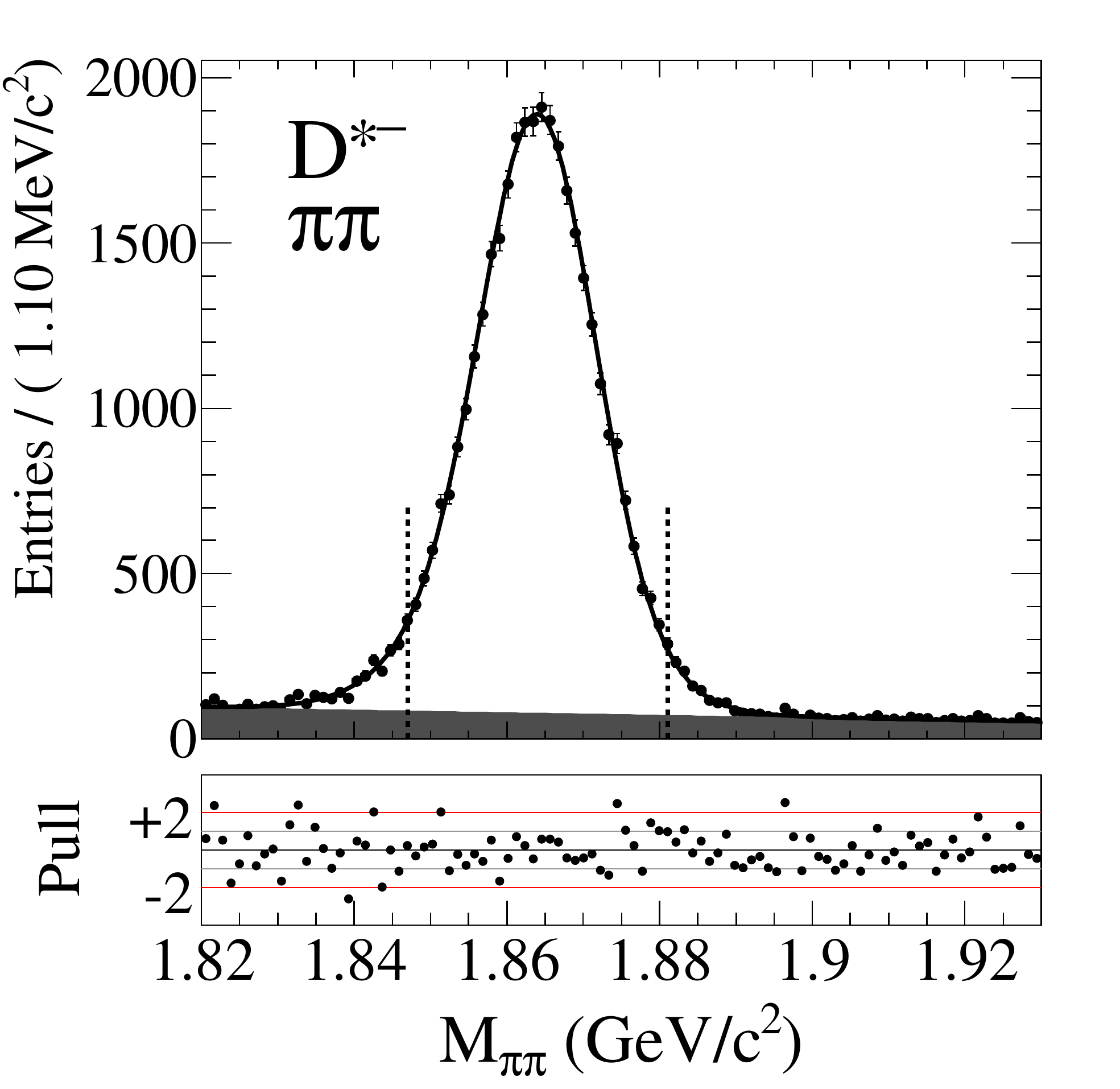}\hfill
  \includegraphics[width=0.24\linewidth]{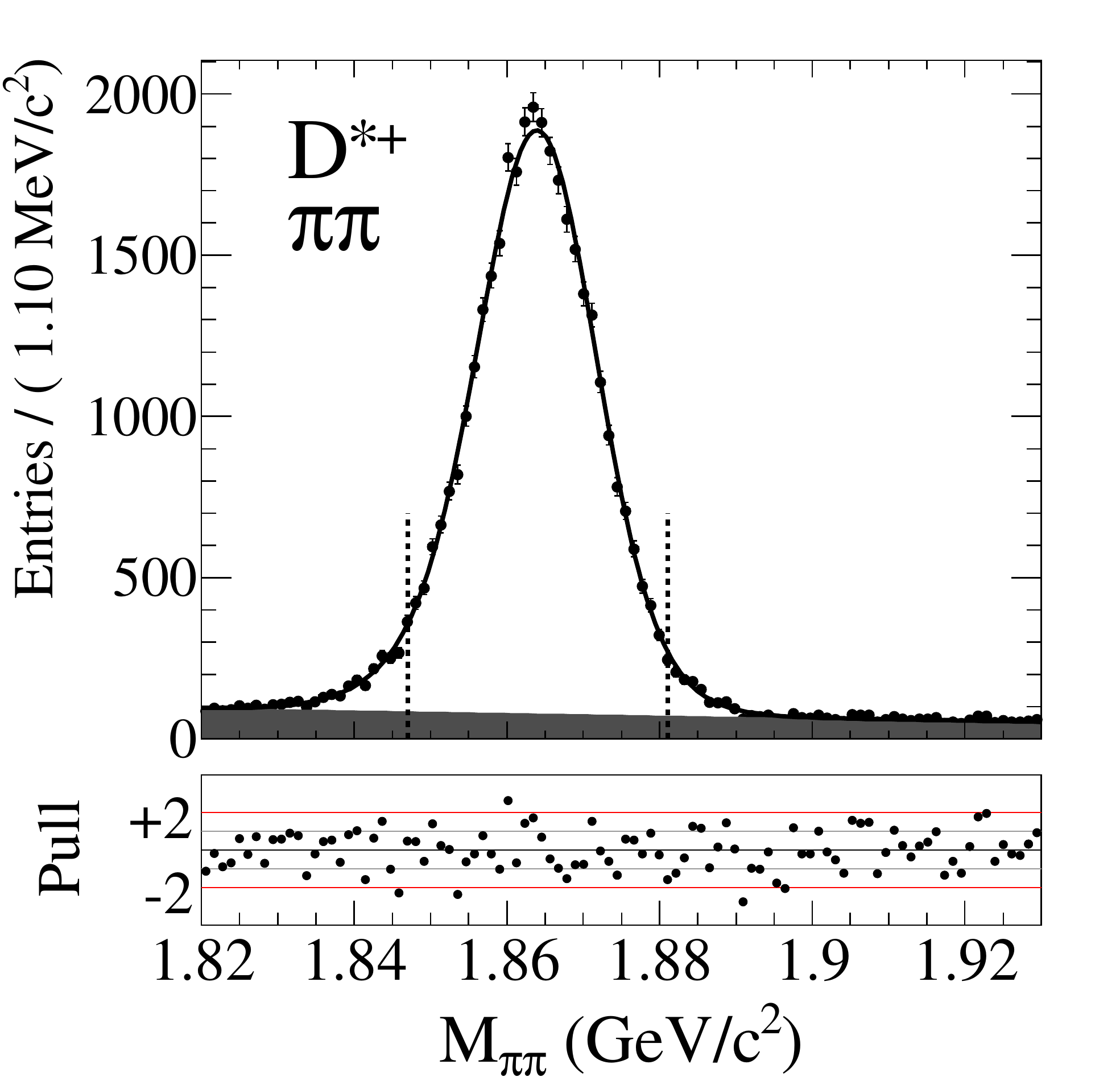}\hfill%\\
  \includegraphics[width=0.24\linewidth]{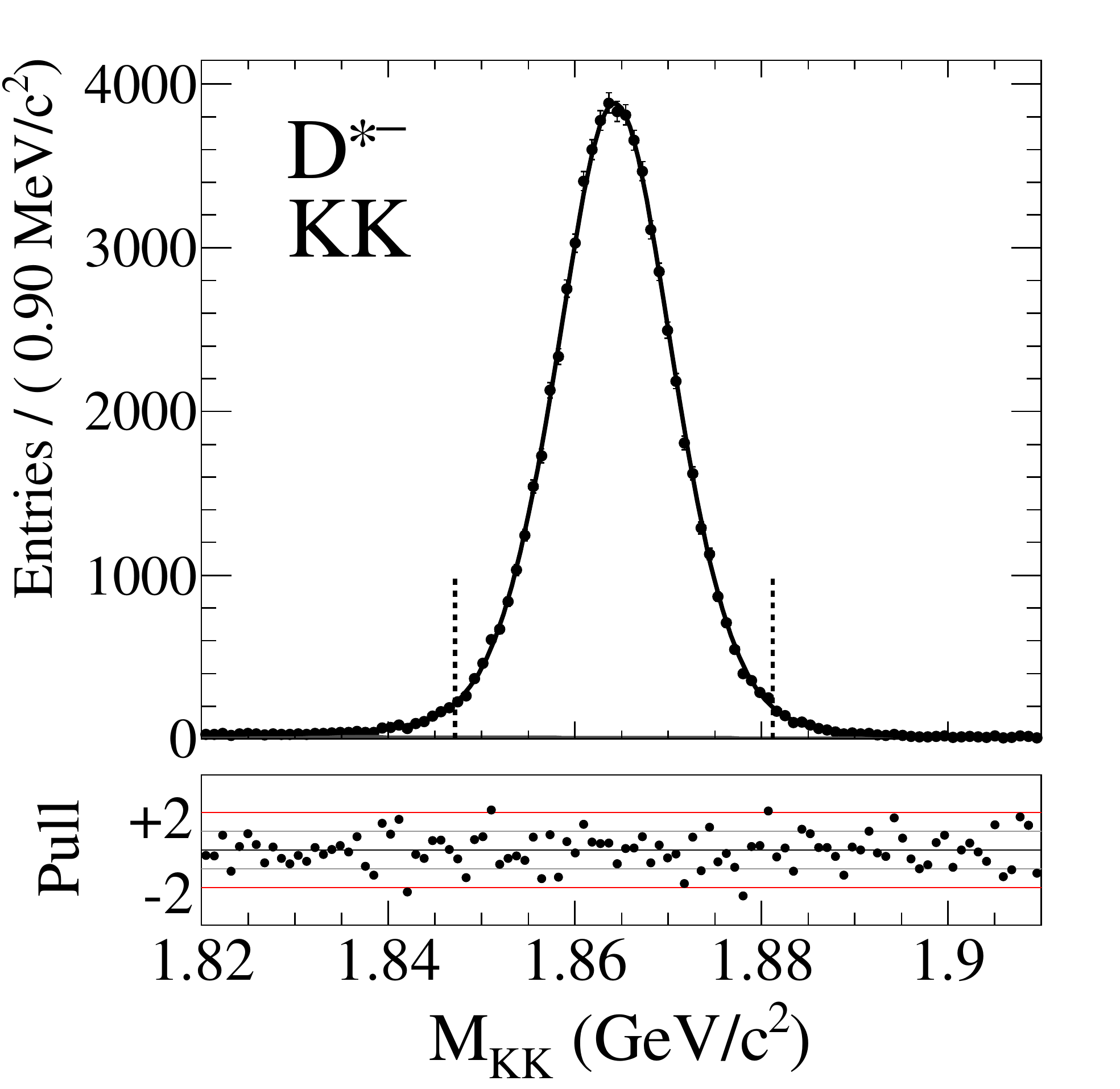}\hfill
  \includegraphics[width=0.24\linewidth]{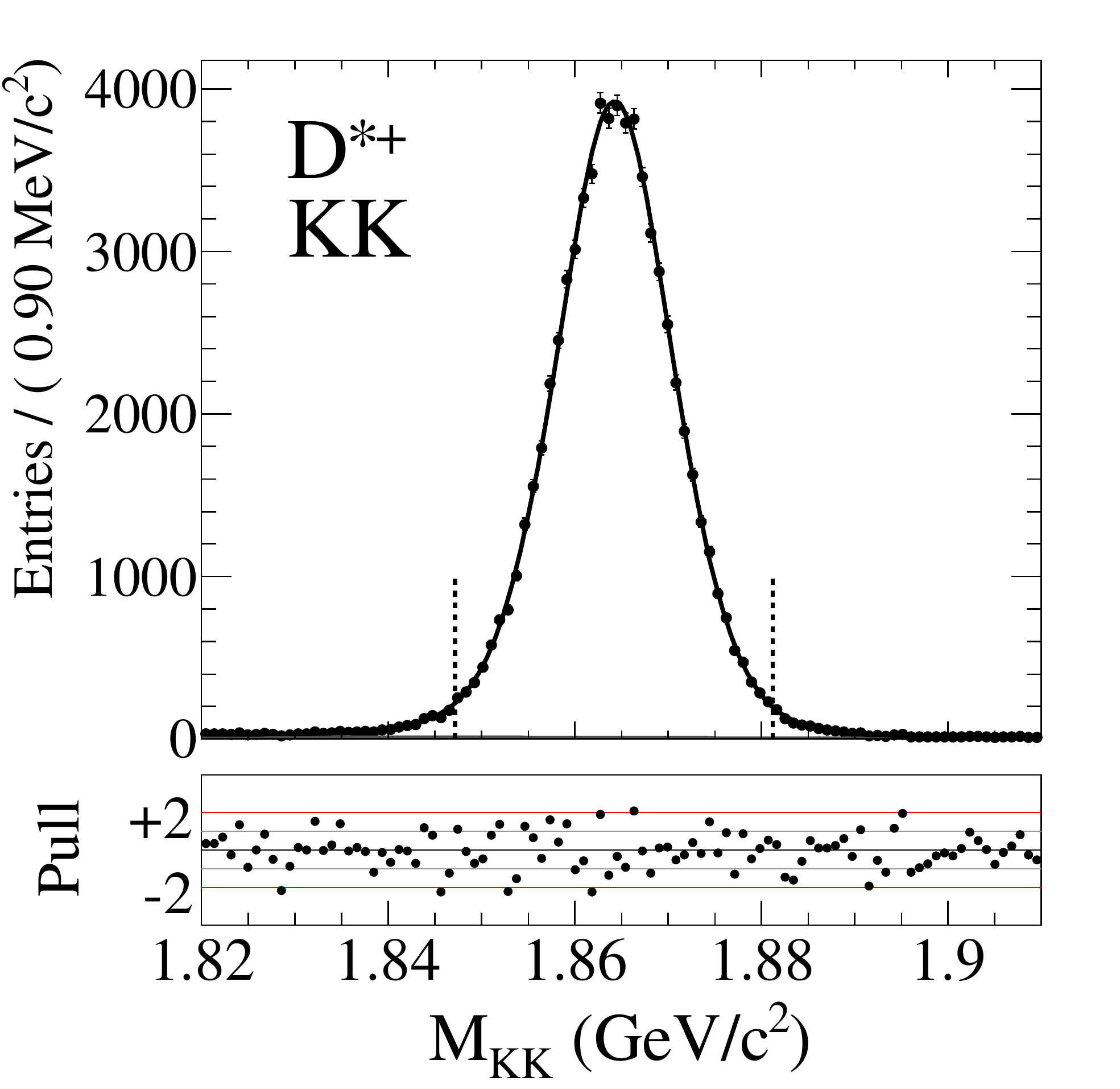}\hfill
}
\hbox to \hsize{\hfill
  \includegraphics[width=0.24\linewidth]{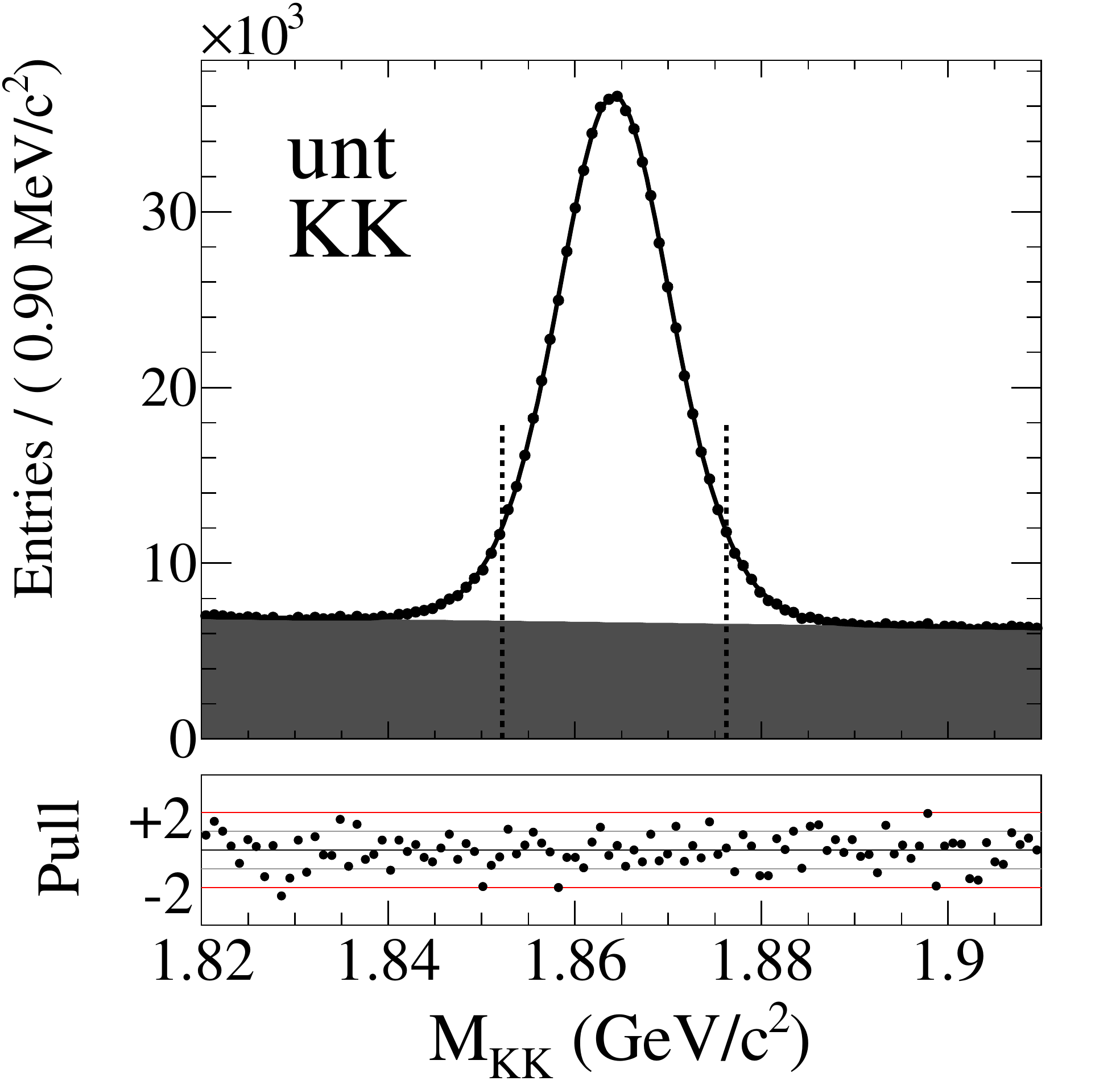}\hfill
 \includegraphics[width=0.24\linewidth]{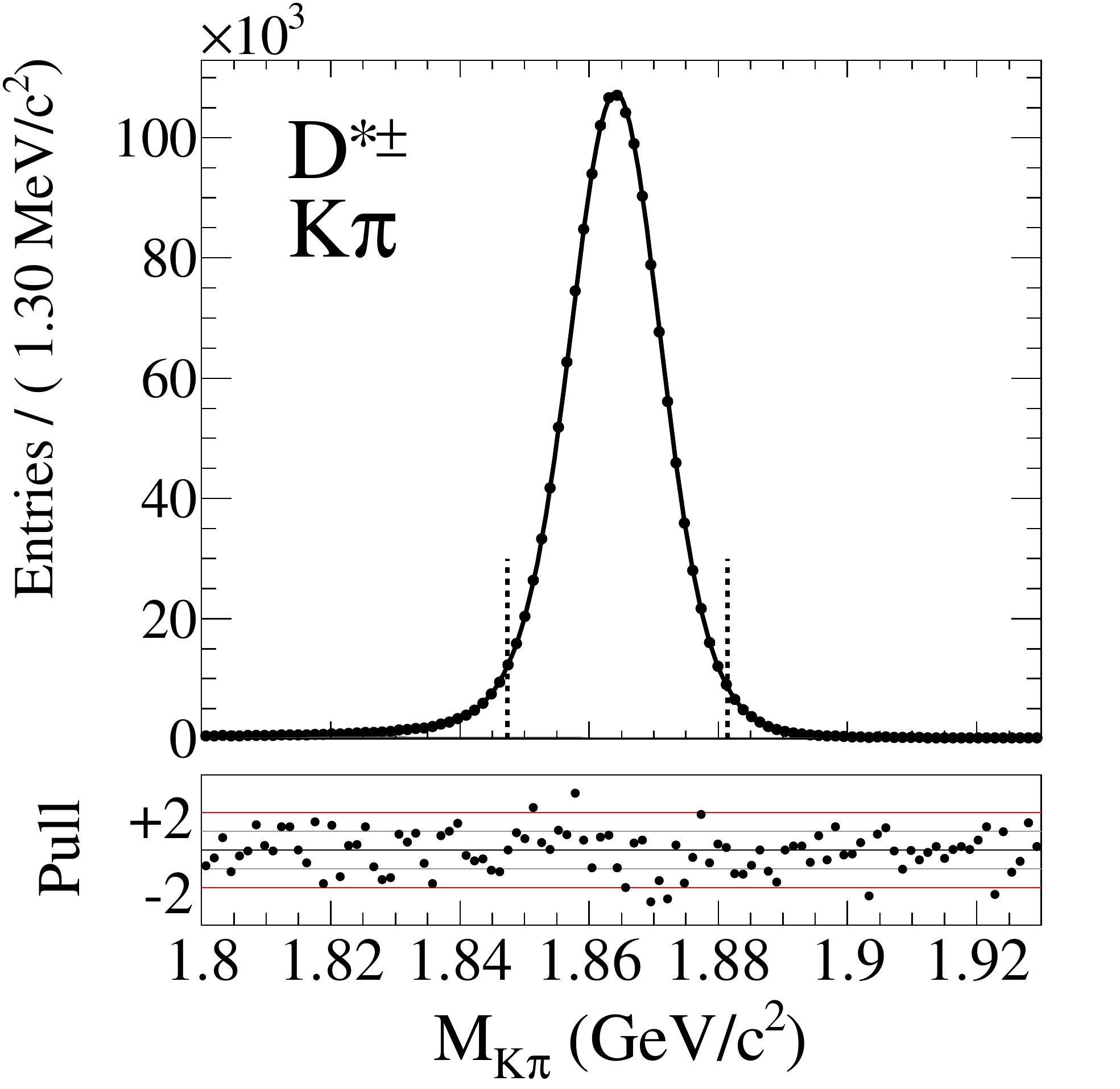}\hfill
 \includegraphics[width=0.24\linewidth]{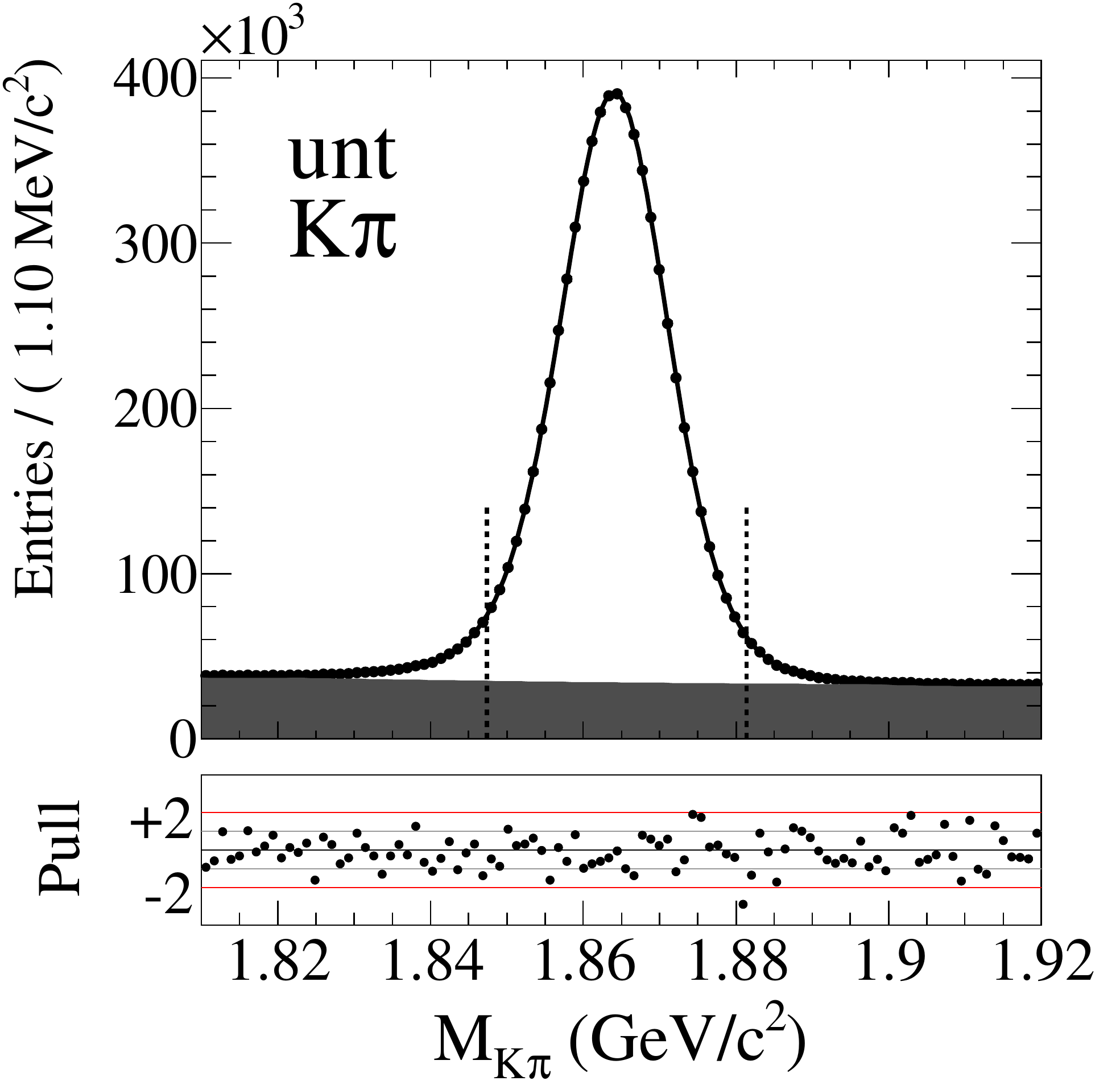}\hfill
}
  \caption{\label{fig:MassPlot}The reconstructed two-body mass distribution
    for the seven modes. The vertical lines show the \emph{signal
    region}. The shaded regions are the background contributions.
    The normalized Poisson residuals for each fit are shown under each plot.}
\end{figure}

In Fig.~\ref{fig:MassPlot} we show the reconstructed invariant mass distributions for the selected \Dz candidates in both tagged and untagged modes.
We fit the mass distributions in order to extract the total number of background candidates.
In Fig.~\ref{fig:MassPlot} we also report the fit results and, below each plot, show the normalized Poisson pulls~\cite{Baker:1983tu}.
For the tagged CP-even modes, the \Dz and \Dzb samples are fit simultaneously, sharing all parameters except for the expected
signal and background candidate yields.

We perform a mode-dependent, data-driven optimization of the invariant mass window position and width, in order to 
reduce significantly the effects of the linear correlation between the reconstructed mass and the reconstructed proper time. The signal regions
obtained for each mode are shown in Fig.~\ref{fig:MassPlot} with dashed lines. 
These are $34\mevcc$ wide for all modes except untagged $\Dz\to\KpKm$. In this mode the signal region width is reduced to $24\mevcc$ due to the higher 
level of background, as observed in the corresponding plot.
We define a lower- and a higher-mass sideband each of width $20 \mevcc$; these are used to study and characterize the combinatorial background.
The mass sidebands for the untagged modes are $\pm 44.5 \mevcc$ from the signal region center.
In case of the tagged modes, the distance of the sidebands from the signal region center is $\pm 35.5 \mevcc$, and the \dm window is shifted
to higher values, $0.151 < \dm < 0.159\gevcc$. In the sidebands the tagged and untagged samples are not independent.

After the selection we divide the backgrounds into two categories. 
Candidates for which the common ancestor of the \Dz products is a long-living charmed meson
are collected in the misreconstructed-charm background category. In Table~\ref{tab:charm} we 
report the composition of this background in the signal region, obtained by studying simulated events.
The other background candidates, consisting mainly of random tracks,
fall into the combinatorial background category. In Table~\ref{tab:yields}
we report the number of signal and background candidates after selection, for the signal region.

\begin{table}[t]
\centering
\begin{tabular}{llllll}
\hline \hline
\multirow{2}{*}{Mode}                    & \multicolumn{3}{c}{Tagged Modes} & \multicolumn{2}{c}{Untagged Modes}     \\ 
                    & \pippim &  \KpKm    &  \DzRW       &  \KpKm         &  \DzRW     \\ 
\hline
$\Dz\to X\ell\nu$   & 15.4      & 10.3       & 29.9      & 7.2       & $\le 2$  \\
$\Dz\to\Km\pip$     & 80.8      & 14.9       & 57.1      & 8.8       & 35.8     \\
$\Dz\to\piz\pip\Km$ & 1.1       & 70.3       & 1.7       & 63.3      & 6.9      \\
$\Dp\to\pip\pip\Km$ & $\le 1$   & 2.9        & $\le 1$   & 11.8      & $\le 2$  \\
$\Dz\to\Kp\Km$      & $\le 1$   & $\le 1$    & 1.3       & $\le 1$   & 3.5      \\
$\Dz\to\pip\pim$    & 1.8       & $\le 1$    & 2.2       & $\le 1$   & 3.1      \\
$\Dz\to\pip\pim\piz$& $\le 1$   & $\le 1$    & 7.0       & $\le 1$   & 17.3     \\
$\Lambda$ decays    & $\le 1$   & $\le 1$    & $\le 1$   & 4.9       & 2.6      \\
\hline\hline
\end{tabular}
\caption{Expected composition of the misreconstructed-charm backgrounds. Only misreconstructed-charm background
modes that have $>1\%$ contribution in at least one signal mode are listed. For the tagged modes,
the fractions are the sum of the separate $\Dz$ and $\Dzb$ tags.}
\label{tab:charm}
\end{table}

\begin{table}[t]
\centering
\begin{tabular}{llll}
\hline \hline
                    & Signal                      & Combinatorial Bkgd.      & Charm Bkgd.    \\ \hline
Tagged \pippim      & $65\,430 \pm 260$           &   $3\,760$               &  97 \\
Tagged \KpKm       & $136\,870 \pm 370$          &   $653$                  & 309  \\
Tagged \DzRW       & $1\,487\,000 \pm 1\,200 $   &   $2\,849$               & 642   \\
Untagged \KpKm      & $496\,200 \pm 1\,200 $      &   $165\,000 \pm 1\,000$  & $5\,477$ \\
Untagged \DzRW      &  $5\,825\,300 \pm 2\,600 $  &   $1\,044\,552$          &$4\,645$  \\
\hline\hline
\end{tabular}
\caption{Signal and background yields in the signal region; yields with
uncertainties are those obtained directly from the lifetime fit to data. For the tagged modes,
the yields are the sum of the separate $\Dz$ and $\Dzb$ tags.}
\label{tab:yields}
\end{table}

\section{Lifetime Fit}
In order to extract the three lifetime values we perform an extended unbinned-maximum-likelihood fit
to the 2-dimensional distribution of proper time and proper time error. All modes are fit simultaneously: the signal resolution
function parameters are shared among the modes, while the background Probability Density Function (PDF) parameters are not.

The single-mode PDF for the signal events consists of an exponential convolved with a 
resolution function. The latter is the sum of three Gaussian functions with a common mean (offset)
and widths proportional to the per-event proper time error, scaled with three different factors (one for each Gaussian).
In order to take into account differences in the reconstruction due to the different final states, we multiply each Gaussian scale
factor by another scale factor that depends on the final state (the \DzRW factor is fixed to 1). In the same way we
introduce a third scale factor that depends whether the mode is tagged or untagged, fixing to unity the one for the untagged 
modes. For the tagged \CP-even modes we also
take into account the wrongly-tagged signal candidates, fixing the fraction of these events to the value 0.2\%, obtained from the simulated events.
Since the proper time PDF depends on the proper time error, we multiply each signal PDF by the 1-dimensional binned distribution of \terr to avoid biases.
The normalization of the proper time PDF is computed for each \terr.
The \terr histogram for the signal events is obtained from the distribution of the events in the signal region after subtraction 
of the misreconstructed-charm and combinatorial-background contributions.

The 2-dimensional PDF for the misreconstructed-charm background is a signal-like PDF, fitted to the simulated events and then fixed in the final fit.
Since this is a physical background, its lifetime, composition and number of events change with the mass window. Therefore we have decided not to
use the sidebands to characterize it.

The combinatorial PDF is determined as a weighted average of the PDFs in the two mass sidebands, which consist of 2-dimensional histograms.
For the untagged \KpKm mode the sideband PDFs for this category are signal-like.
Contributions of signal and misreconstructed-charm in the sidebands are parameterized using the simulated events, and then fixed.
The weighting parameter is determined from simulated events and then is varied as part of the systematic studies.

The expected total-background candidate yields are evaluated from the mass fit and then corrected using the simulated-event information.
The misreconstructed-charm contribution is estimated from the simulated events, and the combinatorial one is obtained by subtraction.
In the final fit the background yields of the two categories are fixed for all modes except for the combinatorial untagged \KpKm mode, where 
it is allowed to float. This became necessary since the prediction of the mass fit was not accurate enough for this mode, where the
combinatorial background represents almost 25\% of the events in the signal region.

\section{Analysis Validation and Systematics}

The validation of the procedure has been performed on four independent samples of simulated events, each equivalent to data integrated luminosity, and also
on a large ensemble of pure Toy MC samples.
We have also performed a qualitative validation on data by running the fit in different configurations. For example, we have fitted the tagged and
untagged samples separately, finding the \KpKm and \DzRW tagged and untagged extracted lifetimes compatible within the statistical uncertainties.
We let the \pippim and \KpKm samples have different lifetimes, allowing for physical effects depending on the final states (direct \CPV and the dependence of $\phi$
on the final state), and found \tauhhp and \tauhhm to be compatible for the two modes.

In addition to these tests, we have also identified sources of systematic error and have evaluated their contributions, as reported
in Table~\ref{tab:SystematicVariations}. We have evaluated the systematic effects due to the choice of the signal region by varying its position and width.
We have varied the fraction of mistagged events in the $\Dz\to h^+h^-$ tagged modes, and the fraction of \Dz's in the untagged \KpKm mode. 
The proper time error PDF is obtained by subtraction of the background distributions. However, in the untagged \KpKm mode, the combinatorial yields are extracted from 
the lifetime fit and not known {\it a priori}. In the nominal fit we first estimate the number of combinatorial events as for the other modes, and use this to perform a first simultaneous fit.
We then repeat the fit using the yields just extracted, and this fit yields the nominal results. In order to evaluate the systematic error associated with this procedure we repeat the fit a third time,
and take as a systematic error estimate the difference from the nominal value. 
We have varied the misreconstructed-charm lifetimes and yields, estimated using the simulated events, by $\gae 2\sigma$.

The combinatorial PDF is extracted from the sidebands after fixing the signal and misreconstructed-charm contributions. We have applied the variations
 described above for the misreconstructed-charm events in the signal region also in the sidebands, for both the signal and the charm-background contributions, and
re-extracted the combinatorial PDF. We have also varied the number of combinatorial-background events in the signal region for the modes in which it was fixed, and the weighting parameter for each mode.

We have varied the selection criteria, in particular that on \terr by $\pm 0.1\ps$. We have also estimated the systematic impact of the best candidate selection by removing or keeping all 
the overlapping candidates. We have estimated the effects of SVT misalignment and have found these to be negligible.

\begin{table}[t]
\begin{center}
\begin{tabular}{l|c|c}  \hline\hline
Fit Variation & $|\Delta[\yCP]|$ (\%)  & $|\Delta[\deltaY]|$ (\%)\\
\hline
mass window width         & 0.057 & 0.022    \\
mass window position      & 0.005 & 0.001    \\
%signal region              & 0.057 & 0.022    \\
\hline
untagged \kk signal \terr PDF         & 0.022 & 0.000     \\
mistag fraction             & 0.000   & 0.000      \\
untagged \kk \Dz fraction & 0.001 & 0.000     \\
%signal description                & 0.022 & 0.000     \\
\hline
charm bkgd. yields        & 0.016 & 0.000      \\
charm bkgd. lifetimes     & 0.042 & 0.001    \\
%charm bkgd.         & 0.045 & 0.001      \\
\hline
comb. yields             & 0.043 & 0.002    \\
comb. sideband weights   & 0.004 & 0.001    \\
comb. PDF shape          & 0.066 & 0.000      \\
%comb. bkgd.               & 0.079 & 0.002      \\
\hline
\terr selection          & 0.052  & 0.053  \\
candidate selection      & 0.028  & 0.011  \\ % should propose to RC to take this out
%  selection               & 0.059  & 0.054  \\ % should propose to RC to take this out

\hline
Total                    & 0.124  & 0.058  \\
%Total                    & 0.12  & 0.06  \\
\hline\hline
\end{tabular}
\caption{The \yCP and \deltaY systematic uncertainty estimates.
The total is the sum-in-quadrature of the entries in each column.}
\label{tab:SystematicVariations}
\end{center}
\end{table}

\section{Results and Conclusions}

\begin{figure}[htb]
%\centering
\hbox to \hsize{
  \includegraphics[width=0.24\linewidth]{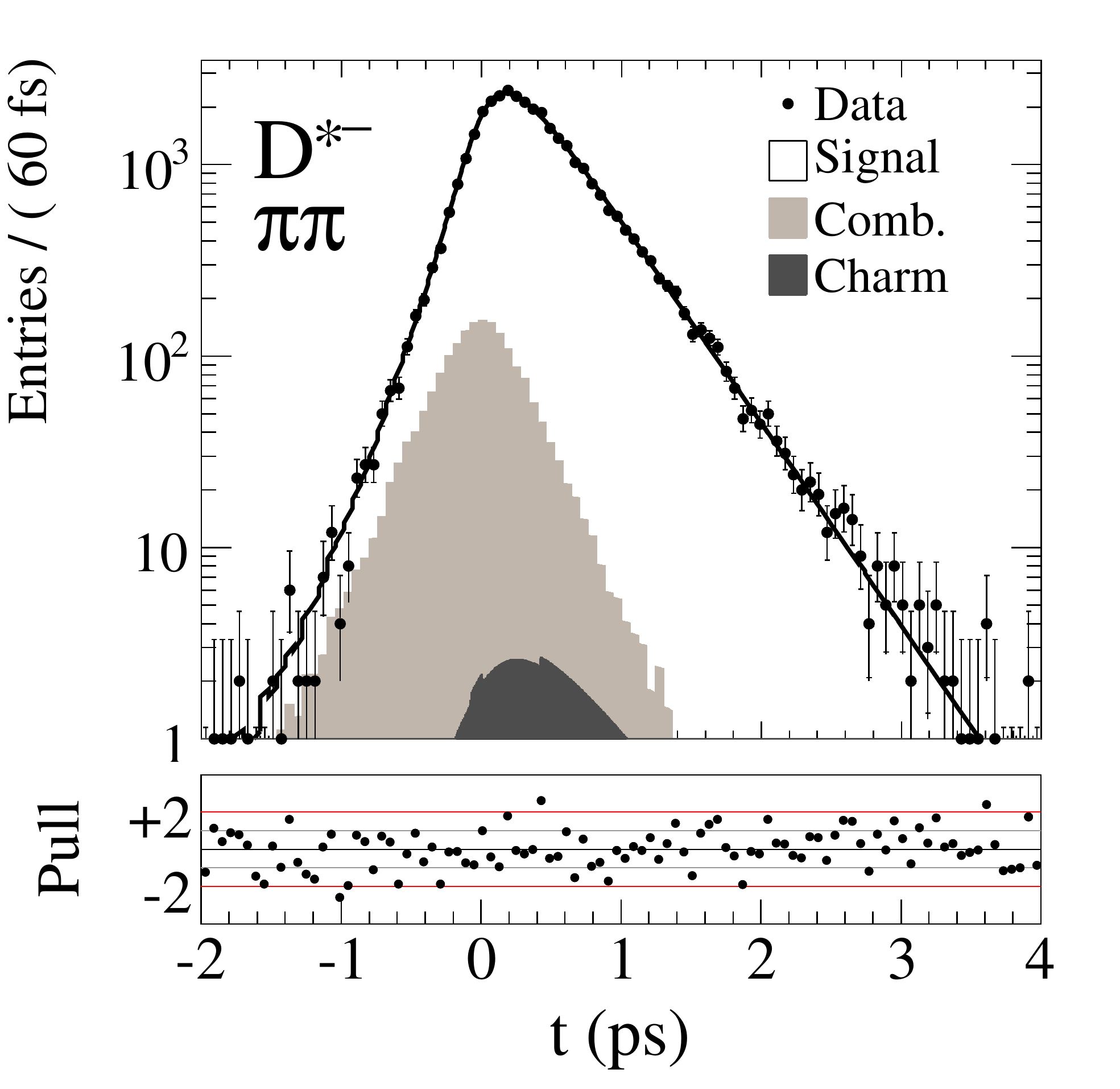}\hfill
  \includegraphics[width=0.24\linewidth]{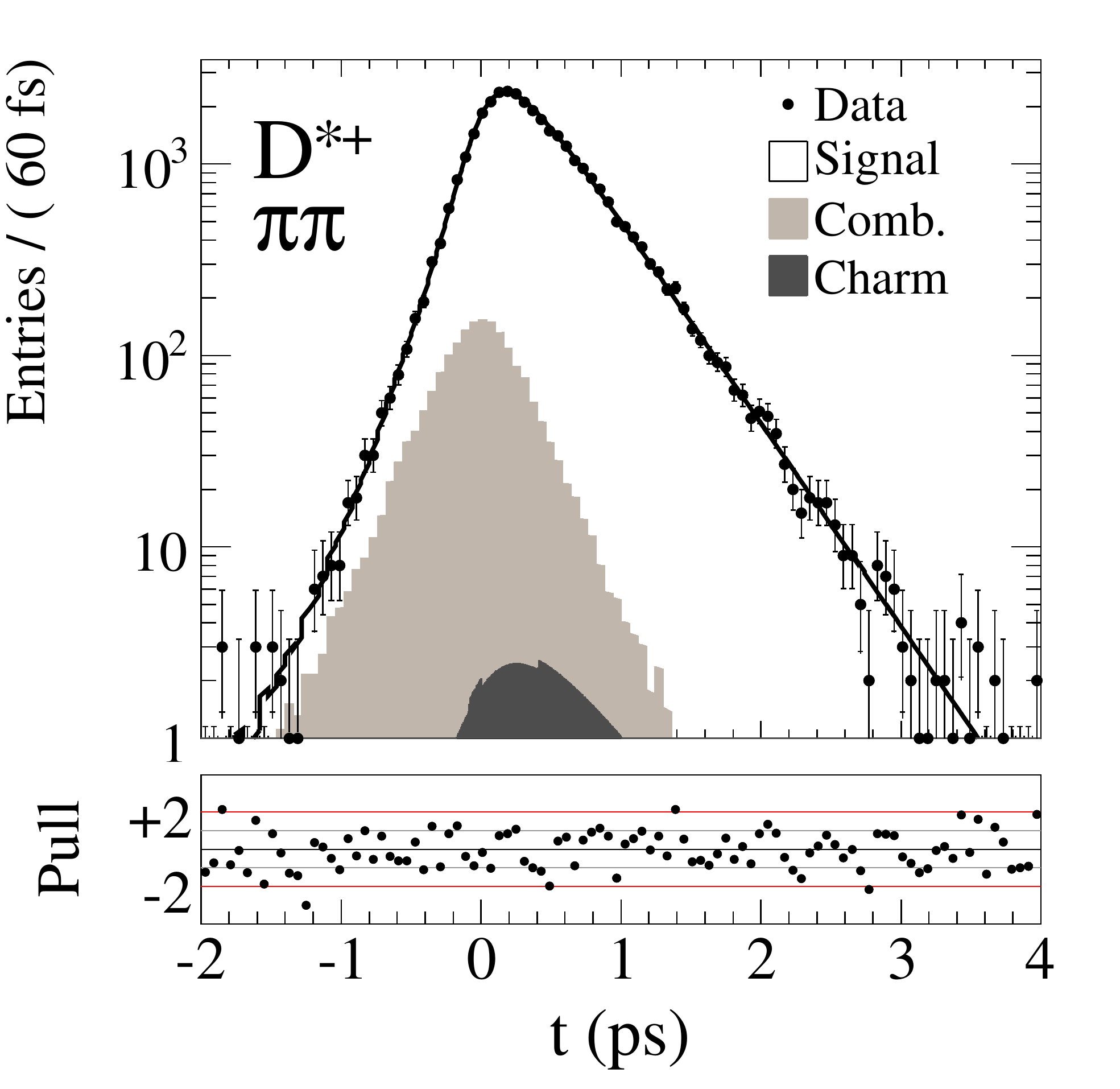}\hfill
  \includegraphics[width=0.24\linewidth]{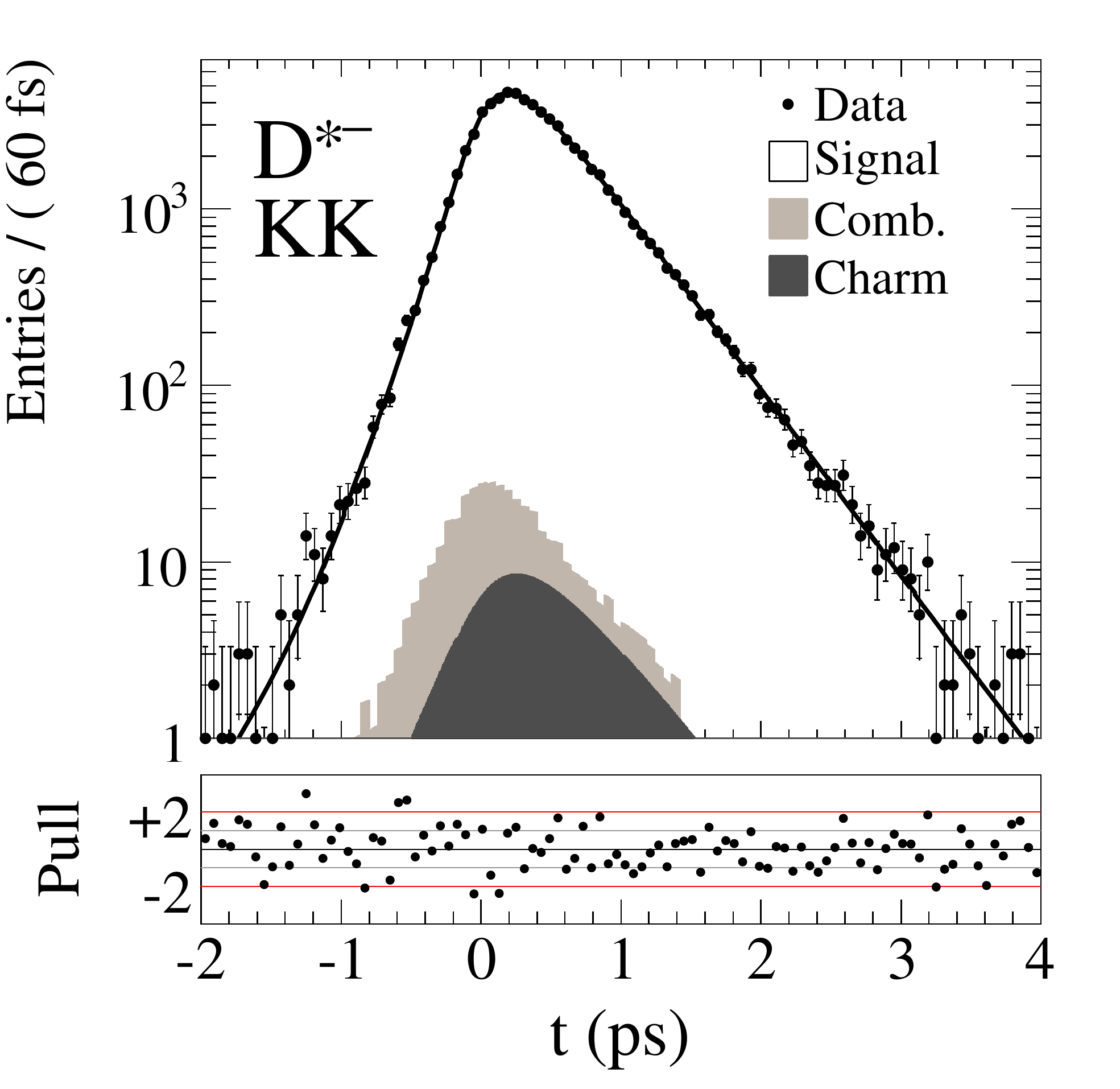}\hfill
  \includegraphics[width=0.24\linewidth]{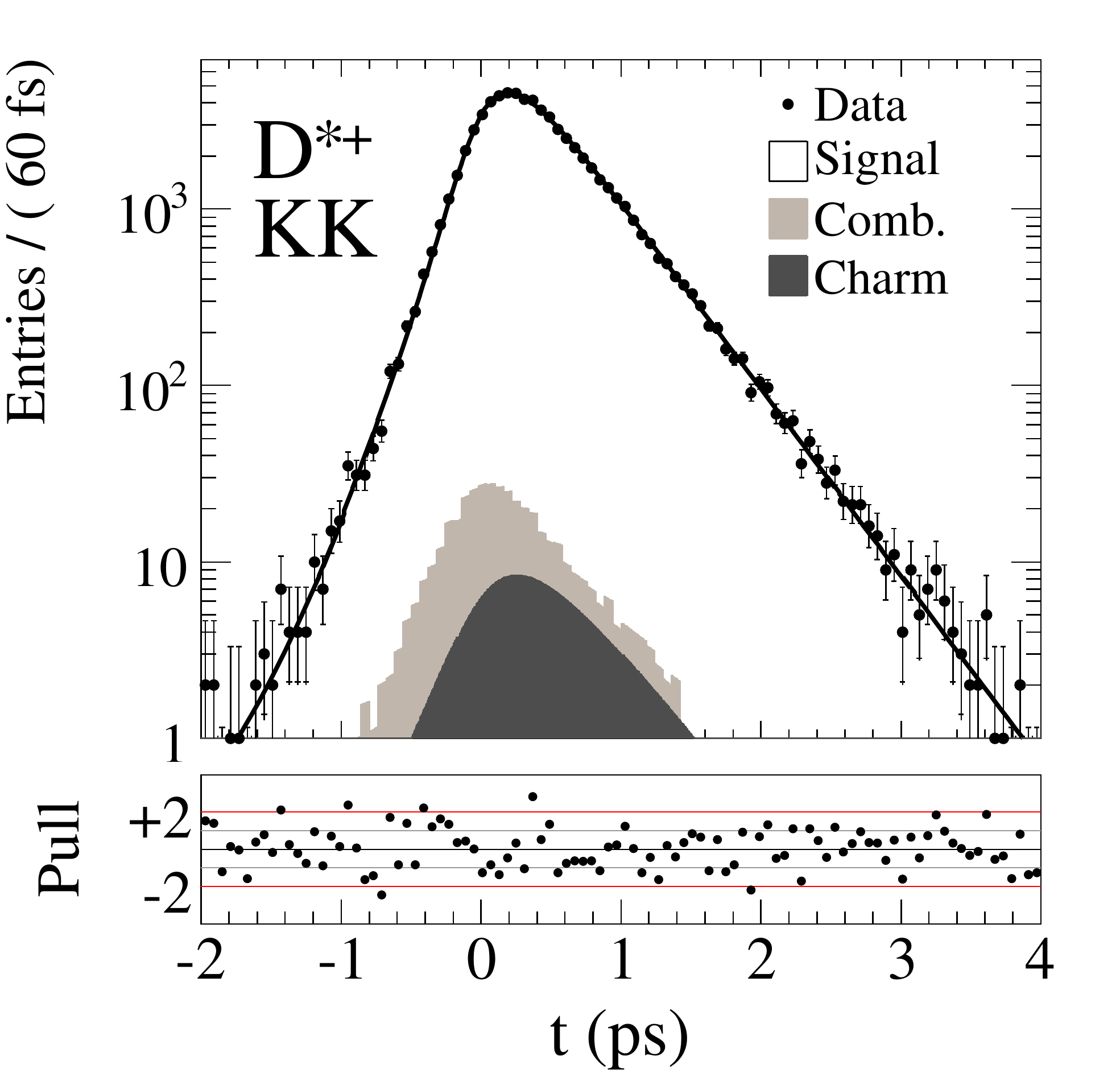}\hfill
}
\hbox to \hsize{
  \includegraphics[width=0.24\linewidth]{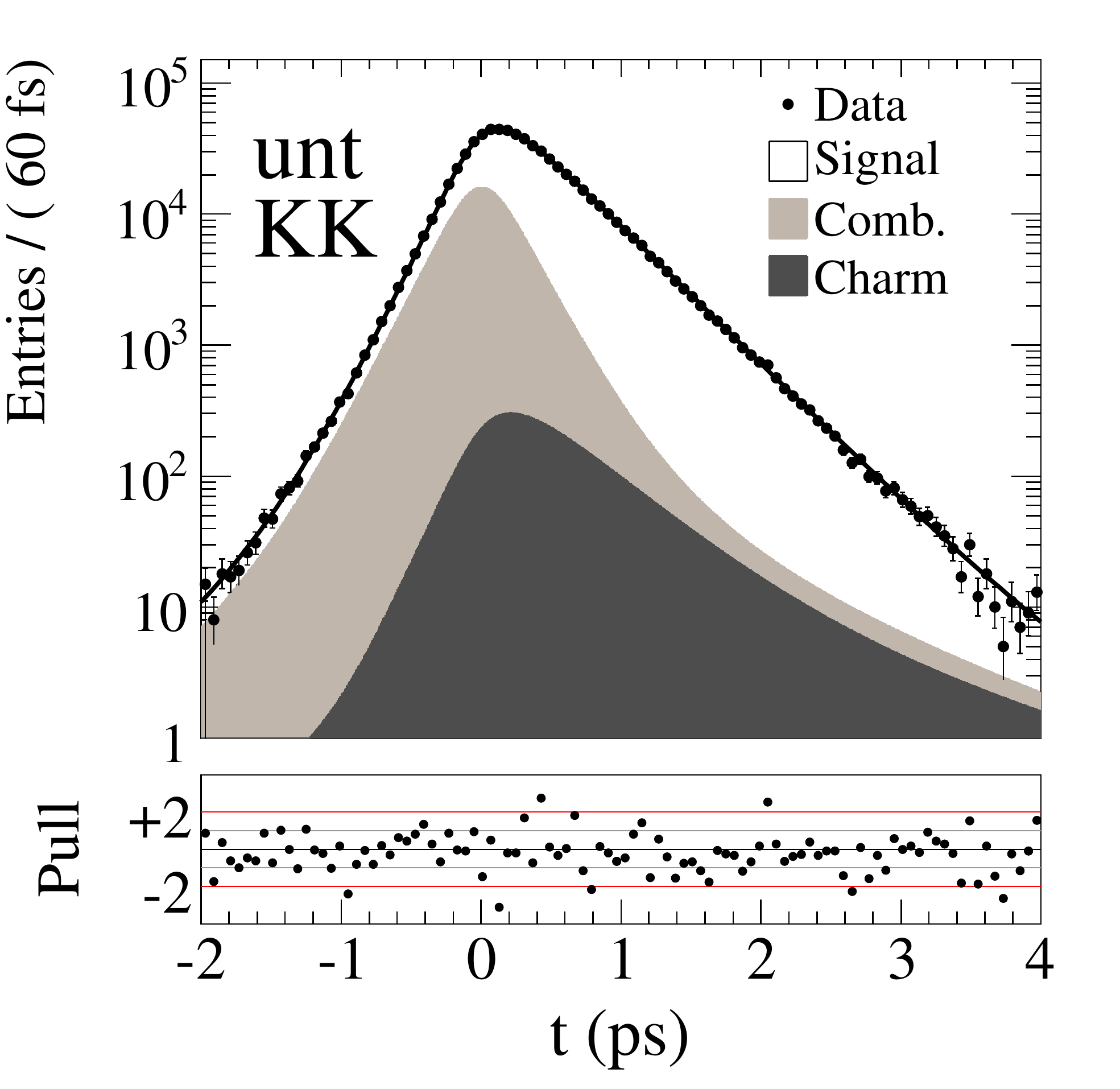}\hfill
 \includegraphics[width=0.24\linewidth]{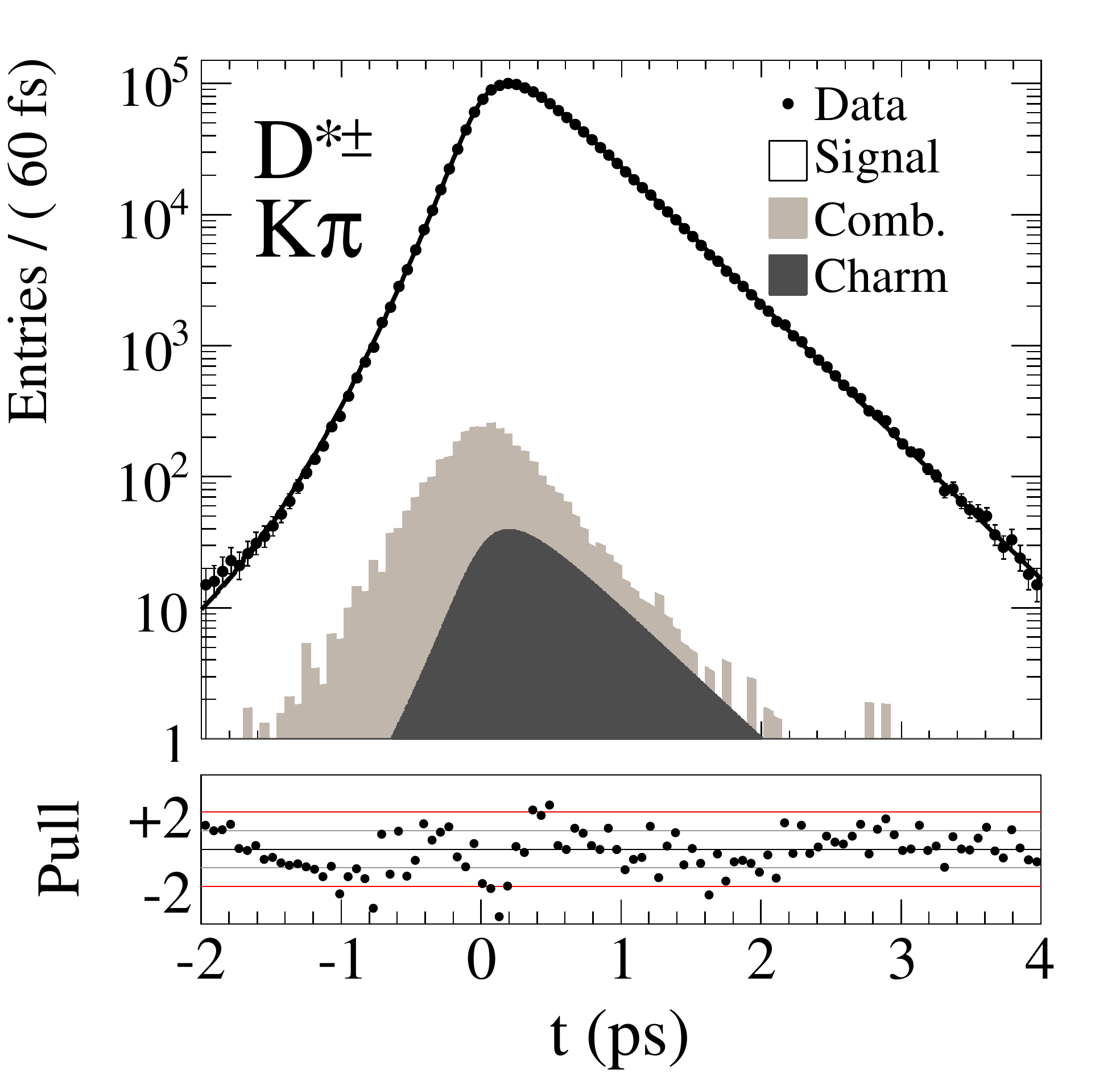}\hfill
 \includegraphics[width=0.24\linewidth]{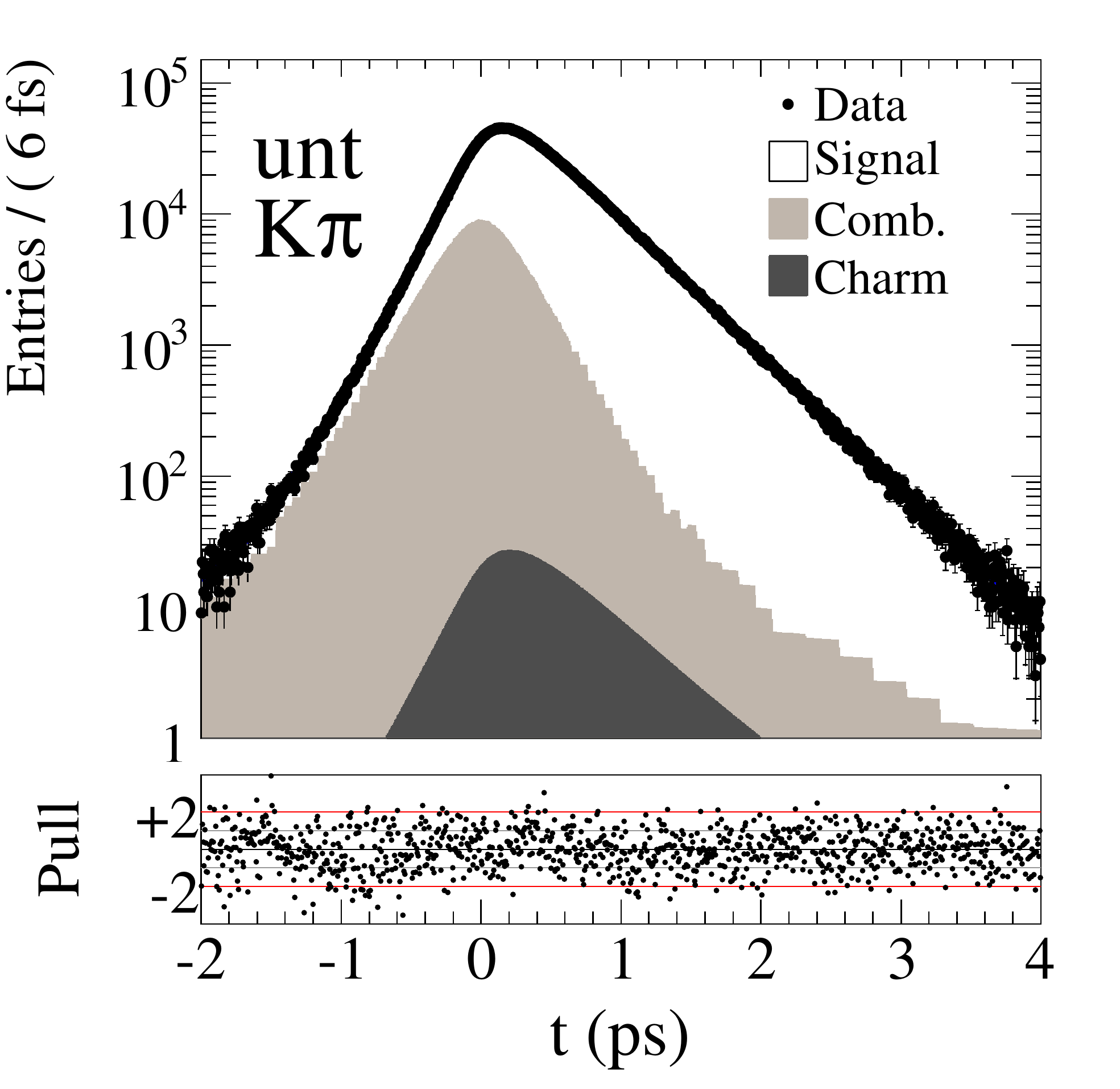}\hfill
 \includegraphics[width=0.24\linewidth]{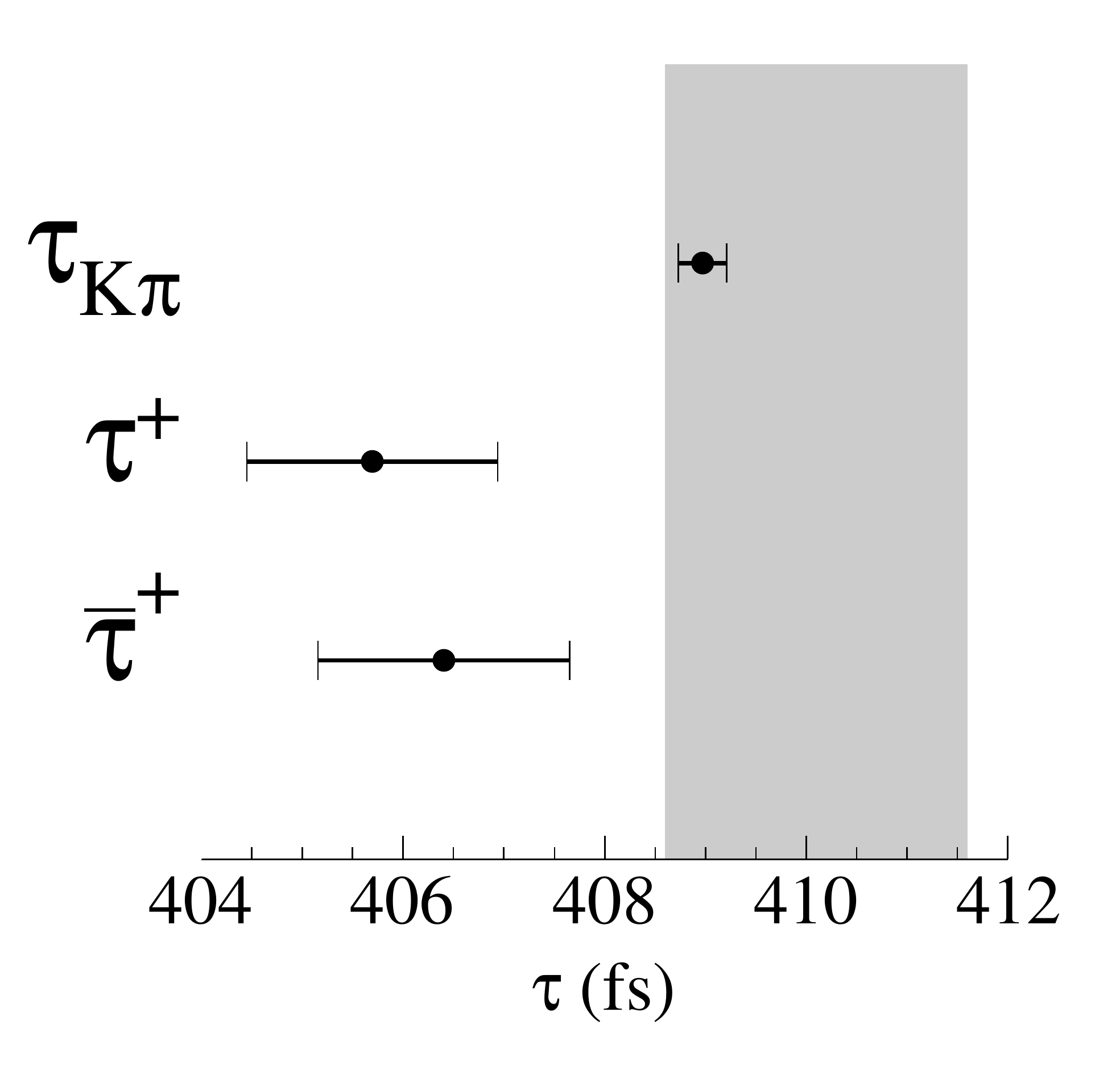}\hfill
}
  \caption{Proper time, $t$, distribution in each final state with the fit result overlaid.
The combinatorial distribution (indicated as 'Comb.' in light gray) is stacked on top of the misreconstructed-charm
distribution (indicated as 'Charm' in dark gray). 
The normalized Poisson pulls for each fit are shown under each plot; ``unt'' indicates the untagged datasets. 
The bottom-right plot shows the individual lifetime values (statistical uncertainty only);
the gray band indicates the PDG \Dz lifetime $\pm1\sigma$~\cite{Nakamura:2010zzi}.\label{fig:timePlot}}
\end{figure}

The seven projections of the lifetime fit are reported in Fig.~\ref{fig:timePlot}.
The following lifetime values are extracted:
\beq
\tauhhp = (405.69\pm1.25)\fs, \qquad \tauhhm = (406.40\pm1.25)\fs, \quad \quad \tauKpi = (408.97\pm0.24)\fs.
\eeq{eq:taus}
The lifetimes are reported with the statistical error only. The \DzRW lifetime is compatible within one standard deviation with
the PDG \Dz lifetime~\cite{Nakamura:2010zzi} and the CP-even lifetimes are significantly lower, as shown in the bottom-right plot of Fig.~\ref{fig:timePlot}.
Combining the values of inverse lifetime following Eq.~(\ref{eq:defycp}), we obtain:
\beq
\yCP = (0.72\pm 0.18 \pm 0.12)\%, \qquad \mbox{and} \qquad \deltaY = (0.09\pm 0.26 \pm 0.09)\fs.
\eeq{eq:obs}
The first error is statistical, obtained from the covariance matrix resulting from the fit, and the second error is systematic.
This measurement represents the most precise measurement of \yCP, and excludes the no-mixing hypothesis at $3.3\sigma$ significance.
The value of \yCP presented here is compatible with all previous measurements. In particular it is compatible  with the previous
 \babar\ measurement~\cite{Aubert:2009ai} with a probability of $\gae 2\%$, taking into account that $\sim40\%$ of
 the events in the current sample are also present in the samples used in the previous measurements~\cite{Aubert:2007en,Aubert:2009ai},
and that the systematic errors are fully correlated. This result favors a lower value for \yCP, and approaches the 
value of the mixing parameter $y$ when measured directly~\cite{Asner:2010qj}, as expected if \CP is conserved.
We find no evidence of \CPV.

In conclusion, we report evidence of \DzDzb mixing with $3.3\sigma$ significance, obtaining the mixing parameter value
 $\yCP =  [0.72 \pm 0.18 \stat \pm 0.12 \syst]\%$. We find no evidence of \CP violation, and measure the \CP-violating parameter value
$\deltaY  =  [0.09 \pm 0.26 \stat \pm 0.06 \syst]\%$.

%%%%%%%%%%%%%%%%%%%%%%%%%%%%%%%%%%%%%%%%%%%%%%%%%%%%%%%%%%%%%%%%%%%%%%%%%
%Recently, F.  A. Mesmer %\index{Mesmer} \index{sensibilities} 
%has reported evidence of a profound influence of 
%magnets on a variety of aspects of human and animal physiology~\cite{Mesmer}.
%reported by Mesmer~\cite{diCenzo,Muller}, but there is little understanding
%%%%%%%%%%%%%%%%%%%%%%%%%%%%%%%%%%%%%%%%%%%%%%%%%%%%%%%%%%%%%%%%%%%%%%%%%

%%%%%%%%%%%%%%%%%%%%%%%%%%%%%%%%%%%%%%%%%%%%%%%%%%%%%%%%%%%%%%%%%%%%%%%%%
% Figure~\ref{fig:magnet}. 
%Table~\ref{tab:blood}.
%%%%%%%%%%%%%%%%%%%%%%%%%%%%%%%%%%%%%%%%%%%%%%%%%%%%%%%%%%%%%%%%%%%%%%%%%

%%%%%%%%%%%%%%%%%%%%%%%%%%%%%%%%%%%%%%%%%%%%%%%%%%%%%%%%%%%%%%%%%%%%%%%%%
%%
%%   use this format to include an  figure into your paper
%%
%\begin{figure}[htb]
%\centering
%\includegraphics[height=1.5in]{magnet}
%\caption{Plan of the magnet used in the mesmeric studies.}
%\label{fig:magnet}
%\end{figure}
%%%%%%%%%%%%%%%%%%%%%%%%%%%%%%%%%%%%%%%%%%%%%%%%%%%%%%%%%%%%%%%%%%%%%%%%%%%

%%%%%%%%%%%%%%%%%%%%%%%%%%%%%%%%%%%%%%%%%%%%%%%%%%%%%%%%%%%%%%%%%%%%%%%%%
%%
%%   use this format to include a LaTeX table  into your paper
%%
%\begin{table}[t]
%\begin{center}
%\begin{tabular}{l|ccc}  
%Patient &  Initial level($\mu$g/cc) &  w. Magnet &  
%w. Magnet and Sound \\ \hline
% Guglielmo B.  &   0.12     &     0.10      &     0.001  \\
% Ferrando di N. &  0.15     &     0.11      &  $< 0.0005$ \\ \hline
%\end{tabular}
%\caption{Blood cyanide levels for the two patients.}
%\label{tab:blood}
%\end{center}
%\end{table}
%%%%%%%%%%%%%%%%%%%%%%%%%%%%%%%%%%%%%%%%%%%%%%%%%%%%%%%%%%%%%%%%%%%%%%%%%%%

\Acknowledgements
% attributions
We are grateful for the excellent luminosity and machine conditions
provided by our \pep2\ colleagues,
and for the substantial dedicated effort from
the computing organizations that support \babar.
The collaborating institutions wish to thank
SLAC for its support and kind hospitality.
This work is supported by
DOE
and NSF (USA),
NSERC (Canada),
CEA and
CNRS-IN2P3
(France),
BMBF and DFG
(Germany),
INFN (Italy),
FOM (The Netherlands),
NFR (Norway),
MES (Russia),
MICIIN (Spain),
STFC (United Kingdom).
Individuals have received support from the
Marie Curie EIF (European Union)
and the A.~P.~Sloan Foundation (USA).

%\bibitem{Mesmer}
%F. A. Mesmer, Proc. Wien. Acad. Sci. {\bf 13}, 1564, 1593 (1762).
%%CITATION = PWASA,13,1564;%%
%\bibitem{diCenzo}
%A. L. di Cenzo, Trans. Acad. Ducal.  Milan., {\bf 23}, 2647 (1771).
%%CITATION = TAADD,23,2647;%%
%\bibitem{Muller}
%For an exhaustive review of the Prussian literature, see A. D. M\"uller,
%in Workshop on Action at a Distance, F. Eisenschmidt, ed. (Springer,
%Berlin, 1774).
%\bibitem{daPonte}
%L. da Ponte, Trans. N. Y. Acad. Sci., {\bf 3}, 27 (1795).
%%CITATION = TNYAA,3,27;%%
 

\begin{thebibliography}{99}

%%
%%  bibliographic items can be constructed using the LaTeX format in SPIRES:
%%    see    http://www.slac.stanford.edu/spires/hep/latex.html
%%  SPIRES will also supply the CITATION line information; please include it.
%%

%\cite{Aubert:2007wf}
\bibitem{Aubert:2007wf}
  B.~Aubert {\it et al.} (\babar\ Collaboration),
  %``Evidence for $D^0$ -anti-D0 Mixing,''
  Phys.\ Rev.\ Lett.\  {\bf 98}, 211802 (2007).
% WS Kpi
%  [arXiv:hep-ex/0703020].
  %%CITATION = PRLTA,98,211802;%%

%\cite{Aubert:2007en}
\bibitem{Aubert:2007en}
  B.~Aubert {\it et al.}  (\babar\ Collaboration),
  %``Measurement of D0 - anti-D0 mixing using the ratio of lifetimes for the
  %decays D0 ---> K- pi+, K- K+, and pi- pi+,''
  Phys.\ Rev.\  D {\bf 78}, 011105 (2008).
%  [arXiv:0712.2249 [hep-ex]].
  %%CITATION = PHRVA,D78,011105;%%

%\cite{Aubert:2009ai}
\bibitem{Aubert:2009ai}
  B.~Aubert {\it et al.}  (\babar\ Collaboration),
  %``Measurement of D0-D0bar Mixing using the Ratio of Lifetimes for the Decays
  %D0->K-pi+ and K+K-,''
  Phys.\ Rev.\  D {\bf 80}, 071103 (2009).
%  [arXiv:0908.0761 [hep-ex]].
  %%CITATION = PHRVA,D80,071103;%%

%\cite{Staric:2007dt}
\bibitem{Staric:2007dt}
  M.~Staric {\it et al.} (Belle Collaboration),
  %``Evidence for $D^0$ - $\bar{D}^0$ Mixing,''
  Phys.\ Rev.\ Lett.\  {\bf 98}, 211803 (2007).
%  [arXiv:hep-ex/0703036].
  %%CITATION = PRLTA,98,211803;%%

%\cite{Abe:2007rd}
\bibitem{Abe:2007rd}
  K.~Abe {\it et al.} (Belle Collaboration),
  %``Measurement of D0-D0bar mixing in D0->Ks pi+ pi- decays,''
  Phys.\ Rev.\ Lett.\  {\bf 99}, 131803 (2007).
%  [arXiv:0704.1000 [hep-ex]].
  %%CITATION = PRLTA,99,131803;%%


%\cite{CDF:2007uc}
\bibitem{CDF:2007uc}
  T.~Aaltonen {\it et al.} (CDF Collaboration),
  %``Evidence for $D^0 - \bar{D}^0$ mixing using the CDF II Detector,''
  Phys.\ Rev.\ Lett.\  {\bf 100}, 121802 (2008).
  %[arXiv:0712.1567 [hep-ex]].
  %%CITATION = PRLTA,100,121802;%%

%\cite{Aaij:2011in}
\bibitem{Aaij:2011in} 
  R.~Aaij {\it et al.}  (LHCb Collaboration),
  %``Evidence for CP violation in time-integrated D0 -> h-h+ decay rates,''
  Phys.\ Rev.\ Lett.\  {\bf 108}, 111602 (2012).
%  [arXiv:1112.0938 [hep-ex]].
  %%CITATION = ARXIV:1112.0938;%%

%\cite{Aaltonen:2011se}
%%bibitem{Aaltonen:2011se}
%  T.~Aaltonen {\it et al.}  (CDF Collaboration),
%%  %``Measurement of CP--violating asymmetries in $D^0\to\pi^+\pi^-$ and $D^0\to K^+K^-$ decays at CDF,''
%  Phys.\ Rev.\ D {\bf 85}, 012009 (2012).
%%  [arXiv:1111.5023 [hep-ex]].
%  %%CITATION = ARXIV:1111.5023;%%

%\cite{Collaboration:2012qw}
\bibitem{Collaboration:2012qw} 
  T.~Aaltonen {\it et al.}  [CDF Collaboration],
  %``Measurement of the difference of CP--violating asymmetries in $D^0 \to K^+K^-$ and $D^0 \to \pi^+\pi^-$ decays at CDF,''
  arXiv:1207.2158 [hep-ex].
  %%CITATION = ARXIV:1207.2158;%
%
%\cite{Liu:1994ea}
\bibitem{Liu:1994ea} 
  T.~-h.~(T.~)Liu,
  %``The d0 anti-D0 mixing search: Current status and future prospects,''
  In *Batavia 1994, The future of high-sensitivity charm experiments* 375-394
  [hep-ph/9408330].
  %%CITATION = HEP-PH/9408330;%%

%\cite{Kagan:2009gb} %used in EPAPS
\bibitem{Kagan:2009gb} 
  A.~L.~Kagan and M.~D.~Sokoloff,
  %``On Indirect CP Violation and Implications for D0 - anti-D0 and B(s) - anti-B(s) mixing,''
  Phys.\ Rev.\ D {\bf 80}, 076008 (2009)
%  [arXiv:0907.3917 [hep-ph]].
  %%CITATION = ARXIV:0907.3917;%%

%%\cite{CC:2008}
\bibitem{CC:2008}
  Charge conjugation is implied throughout.

%\cite{Aubert:2001tu}
\bibitem{Aubert:2001tu}
  B.~Aubert {\it et al.} (\babar\ Collaboration),
  %``The BaBar detector,''
  Nucl.\ Instrum.\ Meth.\  A {\bf 479}, 1 (2002).
%  [arXiv:hep-ex/0105044].
  %%CITATION = NUIMA,A479,1;%%

%\cite{Baker:1983tu}
\bibitem{Baker:1983tu}
  S.~Baker and R.~D.~Cousins,
  %``Clarification Of The Use Of Chi Square And Likelihood Functions In Fits To
  %Histograms,''
  Nucl.\ Instrum.\ Meth.\  {\bf 221}, 437 (1984).
  %%CITATION = NUIMA,221,437;%%

%\cite{Nakamura:2010zzi}
\bibitem{Nakamura:2010zzi}
  K.~Nakamura {\it et al.}  (Particle Data Group),
  %``Review of particle physics,''
  J.\ Phys.\ G {\bf 37}, 075021 (2010).
  %%CITATION = JPHGB,G37,075021;%%

%\cite{Asner:2010qj}
\bibitem{Asner:2010qj}
  D.~Asner {\it et al.}  (HFAG Collaboration),
  %``Averages of b-hadron, c-hadron, and $\tau-lepton Properties,''
  arXiv:1010.1589 [hep-ex].
  %%CITATION = ARXIV:1010.1589;%%


%\cite{Wolfenstein:1985ft}
%\bibitem{Wolfenstein:1985ft}
%  L.~Wolfenstein,
%  %``D0 Anti-D0 Mixing,''
%  Phys.\ Lett.\  B {\bf 164}, 170 (1985).
%  %%CITATION = PHLTA,B164,170;%%

%\cite{Donoghue:1985hh}
%\bibitem{Donoghue:1985hh}
%  J.~F.~Donoghue, E.~Golowich, B.~R.~Holstein, and J.~Trampetic,
%  %``Dispersive Effects In D0 Anti-D0 Mixing,''
%  Phys.\ Rev.\  D {\bf 33}, 179 (1986).
%  %%CITATION = PHRVA,D33,179;%%

%%\cite{Bigi:2000wn}
%\bibitem{Bigi:2000wn}
%  I.~I.~Bigi and N.~G.~Uraltsev,
%  %``D0 anti-D0 oscillations as a probe of quark-hadron duality,''
%  Nucl.\ Phys.\  B {\bf 592}, 92 (2001).
%%  [arXiv:hep-ph/0005089].
%  %%CITATION = NUPHA,B592,92;%%

%\cite{Falk:2001hx}
%\bibitem{Falk:2001hx}
%  A.~F.~Falk, Y.~Grossman, Z.~Ligeti, and A.~A.~Petrov,
%  %``SU(3) breaking and D0 - anti-D0 mixing,''
%  Phys.\ Rev.\  D {\bf 65}, 054034 (2002).
%  [arXiv:hep-ph/0110317].
%  %%CITATION = PHRVA,D65,054034;%%

%%\cite{Falk:2004wg}
%\bibitem{Falk:2004wg}
%  A.~F.~Falk {\it et al.} %, Y.~Grossman, Z.~Ligeti, Y.~Nir, and A.~A.~Petrov,
%  %``The D0 - anti-D0 mass difference from a dispersion relation,''
%  Phys.\ Rev.\  D {\bf 69}, 114021 (2004).
%%  [arXiv:hep-ph/0402204].
%  %%CITATION = PHRVA,D69,114021;%%

%%\cite{Burdman:2003rs}
%\bibitem{Burdman:2003rs}
%  G.~Burdman and I.~Shipsey,
%  %``$D^0$ - $\bar{D}^0$ mixing and rare charm decays,''
%  Ann.\ Rev.\ Nucl.\ Part.\ Sci.\  {\bf 53}, 431 (2003).
%  [arXiv:hep-ph/0310076].
%  %%CITATION = ARNUA,53,431;%%
%
%%\cite{Petrov:2006nc}
%\bibitem{Petrov:2006nc}
%  A.~A.~Petrov,
%  %``Charm mixing in the Standard Model and beyond,''
%  Int.\ J.\ Mod.\ Phys.\  A {\bf 21}, 5686 (2006).
%%  [arXiv:hep-ph/0611361].
%  %%CITATION = IMPAE,A21,5686;%%
%
%%\cite{Golowich:2006gq}
%\bibitem{Golowich:2006gq}
%  E.~Golowich, S.~Pakvasa, and A.~A.~Petrov,
%  %``New physics contributions to the lifetime difference in D0 - anti-D0
%  %mixing,''
%  Phys.\ Rev.\ Lett.\  {\bf 98}, 181801 (2007).
%  [arXiv:hep-ph/0610039].
%  %%CITATION = PRLTA,98,181801;%%
%
%%\cite{Golowich:2007ka}
%\bibitem{Golowich:2007ka}
%  E.~Golowich {\it et al.} %J.~Hewett, S.~Pakvasa, and A.~A.~Petrov,
%  %``Implications of $D^0$ - $\bar{D}^0$ Mixing for New Physics,''
%  Phys.\ Rev.\  D {\bf 76}, 095009 (2007).
%  [arXiv:0705.3650 [hep-ph]].
% %%CITATION = PHRVA,D76,095009;%%
%
%%\cite{Golowich:2009ii}
%\bibitem{Golowich:2009ii}
%  E.~Golowich {\it et al.} %, J.~Hewett, S.~Pakvasa, and A.~A.~Petrov,
%%``Relating $D^0$-$\bar{D}^0$ Mixing and $D^0\rightarrow l^+l^-$ with New Physics,''
%  Phys.\ Rev.\ D {\bf 79}, 114030 (2009).
%%  [arXiv:0903.2830 [hep-ph]].
%  %%CITATION = 0903.2830;%%"
%
%%\cite{Blaylock:1995ay}
%\bibitem{Blaylock:1995ay}
%  G.~Blaylock, A.~Seiden and Y.~Nir,
%  %``The Role of CP violation in D0 anti-D0 mixing,''
%  Phys.\ Lett.\ B {\bf 355}, 555 (1995).
%%  [hep-ph/9504306].
%  %%CITATION = HEP-PH/9504306;%%
%
%%\cite{Isidori:2011qw}
%\bibitem{Isidori:2011qw} 
%  G.~Isidori {\it et al.} %, J.~F.~Kamenik, Z.~Ligeti and G.~Perez,
%  %``Implications of the LHCb Evidence for Charm CP Violation,''
%  Phys.\ Lett.\ B {\bf 711}, 46 (2012).
%%  [arXiv:1111.4987 [hep-ph]].
%  %%CITATION = ARXIV:1111.4987;%%

%%\cite{Hochberg:2011ru}
%\bibitem{Hochberg:2011ru}
%  Y.~Hochberg and Y.~Nir,
%  %``Relating direct CP violation in D decays and the forward-backward asymmetry in $t\bar t$ production,''
%  arXiv:1112.5268 [hep-ph].
%  %%CITATION = ARXIV:1112.5268;%%
%
%%\cite{Cheng:2012wr}
%\bibitem{Cheng:2012wr}
%  H.~-Y.~Cheng and C.~-W.~Chiang,
%  %``Direct CP violation in two-body hadronic charmed meson decays,''
%  Phys.\ Rev.\ D {\bf 85}, 034036 (2012).
%%  [arXiv:1201.0785 [hep-ph]].
%  %%CITATION = ARXIV:1201.0785;%%
%
%%\cite{Giudice:2012qq}
%\bibitem{Giudice:2012qq}
%  G.~F.~Giudice, G.~Isidori and P.~Paradisi,
%  %``Direct CP violation in charm and flavor mixing beyond the SM,''
%  arXiv:1201.6204 [hep-ph].
%  %%CITATION = ARXIV:1201.6204;%%
%
%%\cite{Oreglia:1980cs}
%\bibitem{Oreglia:1980cs}
%  M.~Oreglia (Ph.D. Thesis, Crystal Ball Collaboration), Stanford SLAC-236, UMI-81-08973 [Appendix D] (1980)
%  %%CITATION = SLAC-R-236;%%
%
%%\cite{Agostinelli:2002hh}
%\bibitem{Agostinelli:2002hh} 
%  S.~Agostinelli {\it et al.}  (GEANT4 Collaboration),
%  %``GEANT4: A Simulation toolkit,''
%  Nucl.\ Instrum.\ Meth.\ A {\bf 506}, 250 (2003).
%  %%CITATION = NUIMA,A506,250;%%


\end{thebibliography}
\end{document}